\documentclass{appolb}
\usepackage{epsfig}

\newcommand{\pl}[1]{Phys. Lett. {\bf #1}}

\begin{document}
\title{Nonunitary neutrino mixing matrix
and CP violating neutrino oscillations
\thanks{Presented by J. Gluza 
at the XXV International Conference 
on Theoretical 
Physics ``Particle Physics and Astrophysics in the 
Standard Model and Beyond, Ustro\'n, Poland, September 2001.
}$,\;$\thanks{Supported by the
Polish Committee for Scientific Research under 
Grants No.~2P03B05418 and No.~2P03B04919. M. C. would like to thank the Alexander von Humboldt Foundation for
fellowship.
}}
\author{M. CZAKON 
\address {
Institut f\"ur Theoretische Physik, Universit\"at Karlsruhe, \\
D-76128 Karlsruhe, Germany} \\ \vspace{.5cm}
J. GLUZA AND M. ZRA\L EK
\address {Department of Field Theory and Particle Physics, \\
Institute  of Physics, University of
Silesia \\
Uniwersytecka 4, PL-40-007 Katowice, Poland}
}

\maketitle

\begin{abstract}
In the standard approach to the neutrino oscillations a
unitary  relation
among weak and mass eigenstates of light neutrinos is imposed. 
However, in many extensions of the SM left-handed, 
active  neutrinos mix with additional heavy neutrino states.
Consequences of this additional mixing, driven by experimental
constraints, on the neutrino  oscillations are considered.    
\end{abstract}

\section{Introduction}

At present, 3 light neutrinos, with masses at
the eV or sub-eV scale \cite{ev} are known to exist. However, much heavier
neutrino states ($m_N \ge {\cal{O}} (M_Z/2)$) are not excluded \cite{ex}.
These, due to kinematical reasons do not contribute directly
to the weak neutrino states which can undergo neutrino oscillations.
They influence, however,  neutrino oscillations since they modify the 
neutrino mixing matrix U. Let $U_\nu$ be the full neutrino mixing 
matrix, then the matrix U of dimension $3 \times 3$ constitutes 
the mixing submatrix of light neutrino states $( \nu_e, \nu_\mu ,
\nu_\tau \leftrightarrow \nu_1, \nu_2, \nu_3$ transitions)

\begin{equation}
U_\nu = \left( \matrix{ U & V \cr
                        V' & U'} \right).
\label{gener}
\end{equation}

The submatrix V (of dimension ($n_R \times 3$)) is responsible for the 
mixing of light neutrinos with $n_R$ heavy states ( $ \nu_e, \nu_\mu ,
\nu_\tau \leftrightarrow \nu_4, \ldots ,  \nu_{n_R-3}$ transitions).
The submatrix $U'$ (of dimension ($n_R \times n_R$)) is responsible for 
mixing among heavy states.
In the conventional see-saw mechanism $m_N >> m_\nu$, where 
$m_N$ and $m_\nu$ are masses of heavy and light states respectively,
the elements of V are very small and U becomes unitary. 
This simply means that heavy neutrino
states do not modify mixings among light neutrino states.
From the
theoretical point of view, V must not necesserily  be negligible \cite{th}.
We will use the experimental data to  constrain
V \cite{exp}, and more precisely the combination 
$\left( VV^\dagger \right)_{\alpha \beta}$\footnote{The
elements of the V matrix can  also be investigated, e.g. in heavy
neutrino production processes \cite{pr}.} $(\alpha, \beta = \{
e, \mu, \tau \})$.

From the unitarity of  $U_\nu$ we infer that
\begin{equation}
\left( UU^\dagger \right)_{\alpha \beta}= \delta_{\alpha \beta}
- \left( VV^\dagger \right)_{\alpha \beta}.
\label{un}
\end{equation}

The aim of this paper is to examine the effect of this modification of unitarity of V on neutrino 
oscillations. 
The subject is not new \cite{lan}\footnote{Lately, effects of a non-unitary mixing 
matrix U have been considered in \cite{gon} in a different context
where new leptonic interactions have been included.}. Nevertheless,
some issues, especially connected with CP-violation effects have not yet
been discussed.
CP violation  effects in the unitary neutrino oscillations 
case are known to be very fragile. If any element of the unitary 
U matrix (e.g. $U_{e3}$) is small then the effect of CP violation
will be small either.
And in fact, $U_{e3}$ (see Eq.~\ref{u}) is known to be very small
if not zero. Besides, the CP phase $\sin \delta$ must be substantial. 
Finally, the CP violating effects vanish with decreasing $\delta
m^2_{\odot}$. For $\delta=
\frac{\pi}{2}$,  $\delta m^2_{\odot}$ given by LMA MSW solution
and $U_{e3} >0$, the CP effects can be detectable 
\cite{cpdet}, but even then it may happen that matter effects
will mimic (or screen) the CP violation \cite{cpmat}. We show that the
nonunitarity
of U can be responsible for similar effects. 
If CP-violation effects were detected
with a strength larger than predicted 
by the  unitary neutrino  mixing
approach, then heavy neutrino mixing could be held responsible
for this effect.
In the contrary case, some better bounds on the 
$\left( VV^\dagger \right)_{\alpha \beta}$ factors could be found.

In this paper we focus on neutrino oscillations in vacuum.

\section{Neutrino oscillations in the presence of heavy 
neutrino states}

 In the standard neutrino oscillation theory of three flavours 
we start with neutrino weak eigenstates
$\nu_\alpha=(\nu_e,\nu_\mu,\nu_\tau)$ as a  combination of 
three mass eigenstates $\nu_i=(\nu_1,\nu_2,\nu_3)$
\begin{equation}
\nu_\alpha= \sum\limits_{i=1}^3 U_{\alpha i} \nu_i .
\label{miesz33}
\end{equation}

The form of the matrix U can be obtained using subsequent
rotations around the axes spanned by massive neutrino states
$m_1,m_2,m_3$
\begin{equation}
U=R_{23} R_{13} R_{12}.
\label{r33}
\end{equation}

$R_{ij}$'s represent rotations in the i-j plane by 
$\Theta_{ij}$ angle with additional phases,
e.g. ($c_{12} \equiv \cos{\Theta_{12}}$, $s_{12} \equiv
 \sin{\Theta_{12}} e^{i \delta_{12}}$)

\begin{equation}
R_{12}=\left( \matrix{c_{12} & s^{\ast}_{12} & 0 \cr
                     -s_{12} & c_{12} & 0 \cr
                       0 & 0 & 1} \right).
\end{equation}

Taking $\delta_{12}=\delta_{23}=0$ (two of the three complex phases 
do not influence the oscillation probability 
 \cite{dow}) we obtain the classical parametrization of the U matrix \cite{pdg}
 ($\delta_{13} \equiv \delta$)
\begin{eqnarray}
U&=&\left( 
\begin{array}{ccc}
U_{e1} & U_{e2} & U_{e3} \\ 
U_{\mu 1} & U_{\mu 2} & U_{\mu 3} \\ 
U_{\tau 1} & U_{\tau 2} & U_{\tau 3} 
\end{array}
\right) \nonumber \\
&=&\left( 
\begin{array}{ccc}
c_{12}c_{13} & c_{13} s_{12} & s_{13} e^{-i \delta} \\
           -c_{23}s_{12}-s_{13}s_{23}c_{12}e^{i \delta} 
&  c_{12}c_{23}-s_{12}s_{23}s_{13}e^{i \delta}  & c_{13} s_{23} \\
   s_{12}s_{23}-s_{13}c_{23}c_{12}e^{i \delta} &  
-s_{23}c_{12}-s_{12}c_{23}s_{13}e^{i \delta} & c_{23}c_{13} 
\end{array}
\right) 
\label{u}
\end{eqnarray}

Let us now include effects of the matrix V
to the matrix U (Eqs.~\ref{gener},\ref{un}). We will do it by 
introducing three new
parameters $\epsilon_i$, i=1,2,3 which are 
connected directly to the elements of the matrix V in the case
of the $4 \times 4$ matrix Eq.~\ref{gener}.

The general $4 \times 4$ matrix Eq.~\ref{gener} can be
parametrized by 6 rotation angles (and 6 phases)
in the following way

\begin{equation}
U_\nu=R_{34} R_{24} R_{14} R_{23} R_{13} R_{12},
\label{r44}
\end{equation}

where the rotations take place in the 4 dimensional space spanned
by four massive neutrino states, e.g.  \cite{bar44},\cite{frgener}
($s_{12} \equiv \sin{\Theta_{12}}e^{i \delta_{12}}$)
\begin{equation}
R_{12}=\left( \matrix{c_{12} & s^{\ast}_{12} & 0 & 0\cr
                     -s_{12} & c_{12} & 0 & 0 \cr
                       0 & 0 & 1& 0 \cr 0& 0& 0& 1} \right).
\end{equation}

Let us note that Eq.~\ref{r44} differs from Eq.~\ref{r33}
by three additional rotations $R_{34} R_{24} R_{14}$
in the plane to which the additional fourth neutrino 
state belongs. When the fourth state is much heavier
than the light states, the rotation angles are small. Let us take
$|s_{14}| \equiv |\epsilon_1| <<1$,
$|s_{24}| \equiv |\epsilon_2| <<1$ and
$|s_{34}| \equiv |\epsilon_3| <<1$. Then  
we can expand Eq.~\ref{r44} to get

\begin{equation}
U_\nu=\left( 
\begin{array}{cc}
\biggl( U( \epsilon_i ) \biggr) &
\begin{array}{c}
 \epsilon_1 \\
 \epsilon_2 \\
 \epsilon_3 
\end{array} \\
& \\
 g ( \epsilon_i) & 1-\frac{1}{2}( | \epsilon_1|^2
 +| \epsilon_2|^2+| \epsilon_3| ^2)
\end{array}
\right)
\label{mod}
\end{equation}

In the limit $\epsilon_i \to 0$ $U( \epsilon_i ) \to U$ 
(Eq.~\ref{u}) and $g(  \epsilon_i ) \to 0$.
 $U( \epsilon_i )$ is the desired matrix which we will
use in the neutrino oscillation formula instead of the U matrix
in Eq.~\ref{miesz33}.
We will not show the manifest form of   $U( \epsilon_i )$
as it is straithorward but space consuming. Our parametrization
through (complex) $\epsilon$ factors holds in the general case of n heavy states
and can be easily connected to the quantities which are usualy
constrained by experimental data, e.g.

\begin{equation}
|\left( VV^\dagger \right)_{e e}|= \sum\limits_{i=heavy}
|V_{e i}|^2 \equiv |\epsilon_1|^2 
\leq 0.0054,
\label{ee}
\end{equation}

and similarly,
 
\begin{eqnarray}
| \left( VV^\dagger \right)_{e \mu} | &=& | \epsilon_1 \epsilon_2|
\leq 10^{-4},
\label{emu} \\
| \left( VV^\dagger \right)_{\mu \tau} | &=& | \epsilon_2 \epsilon_3|
\leq 10^{-2}.
\label{mutau} 
\end{eqnarray}

The very strict constraint, Eq.~\ref{emu} 
comes from the lack of the $\mu \to
e \gamma $ decay \cite{exp,lan,chang}. 
Constraints,  Eqs.~\ref{ee},\ref{mutau}
are consequences of global fits to experimental data \cite{exp,lan}
(e.g. lepton universality, invisible Z decay, CKM unitarity).
There is also a constraint on $\sum\limits_{i=heavy}
|V_{e i}|^2/M_i$  coming from the neutrinoless
double beta decay \cite{b00}. In our approach we do not have to use any information
on the  heavy neutrino mass spectrum. We just assume that the masses are above
100 GeV. The constraint from $(\beta \beta)_{0 \nu}$ is then fulfilled.
 With the  parametrization Eq.~\ref{mod} it is 
straightforward to write the  modified neutrino oscillation 
probability 

\begin{eqnarray}
P_{\nu_\alpha \to \nu_\beta} &=& N_\alpha^2 N_\beta^2 \left\{
\left(  \delta_{\alpha \beta}- \left| \left( VV^{\dagger}
\right)_{\alpha \beta} \right| \right)^2 \right. \nonumber \\
&& \nonumber \\
&-& 4 \sum\limits_{a>b} \tilde{R}^{ab}_{\alpha \beta}
\sin^2{\Delta_{ab}}- 8 \tilde{I}^{12}_{\alpha \beta} 
\sin{\Delta_{21}} \sin{\Delta_{31}} \sin{\Delta_{32}} \nonumber \\
&& \nonumber \\
&-&2 \left[  A^{(1)}_{\alpha \beta} \sin{2 \Delta_{31}}+
A^{(2)}_{\alpha \beta} \sin{2 \Delta_{32}} \right] \Biggr\},
\label{prob}
\end{eqnarray}
where
\begin{eqnarray}
\Delta_{ab}&=&1.27 \delta m^2_{ab} [eV] \frac{L [km]}{E [GeV]}, 
\;\;\;\;\;
\delta m^2_{ab} =m_a^2-m_b^2.  
\end{eqnarray}

$ \tilde{R}^{ab}_{\alpha \beta}$  are modified 
definitions taken from  the standard, unitary approach

\begin{eqnarray}
\tilde{R}^{ab}_{\alpha \beta}&=&Re 
\left[ W_{\alpha \beta}^{ab} \right] , \label{re} \\
\tilde{I}^{ab}_{\alpha \beta}&=&Im 
\left[ W_{\alpha \beta}^{ab} \right] , \label{im} \\
W_{\alpha \beta}^{ab}(\epsilon_i)&=&U_{\alpha a} U_{\beta b} U_{\alpha b}^{\ast}  U_{\beta a}^{\ast} \label{w}. 
\end{eqnarray}

$N_{\alpha ( \beta )}$ 
are factors which normalize the three light neutrino states to 1

\begin{eqnarray}
N_\alpha^2 &=& \frac{1}{1- \left( VV^\dagger
\right)_{\alpha \alpha}}.
\end{eqnarray}

The last row in Eq.~\ref{prob} with

\begin{equation}
A_{\alpha \beta}^{(i)}(\epsilon_i) =
Im \left[ U_{\alpha i}^\ast U_{\beta i}
\left( VV^\dagger
\right)_{\alpha \beta} \right] 
\label{ai}
\end{equation}
deserves an extra comment. Its appearence is a consequence
of the modification of the  Jarlskog factors, which for unitary U fullfil
the following relations

\begin{equation}
I^{ab}_{\alpha \beta}=-I^{ba}_{\alpha \beta}=-I^{ab}_{ \beta \alpha}=
I^{ba}_{\beta \alpha}. 
\end{equation}

When U is not unitary, Eq.~\ref{un} 
leads to 

\begin{eqnarray}
\tilde{I}^{12}_{\alpha \beta} & = & - \tilde{I}^{32}_{\alpha \beta}
- Im \left( U^{\ast}_{\alpha 2} U_{\beta 2} \left( V V^{\dagger}
\right)_{\alpha \beta} \right) \\
\tilde{I}^{21}_{\alpha \beta} & = & - \tilde{I}^{31}_{\alpha \beta}
- Im \left( U^{\ast}_{\alpha 1} U_{\beta 1} \left( V V^{\dagger}
\right)_{\alpha \beta} \right) \\
\tilde{I}^{23}_{\alpha \beta} & = & - \tilde{I}^{13}_{\alpha \beta}
- Im \left( U^{\ast}_{\alpha 3} U_{\beta 3} \left( V V^{\dagger}
\right)_{\alpha \beta} \right) 
\end{eqnarray}

and therefore to the last term in Eq.~\ref{prob}.

As discussed in \cite{lan}, the effects of the 
normalization factors will be difficult to observe in experiments.
Here we will focus on the influence of the additional neutrino 
mixing of light neutrinos 
represented by the $\epsilon_i$'s on the CP violating effects.   
The novelty here is the appearance of the third line in 
Eq.~\ref{prob}. This term is not very sensitive to 
$\Delta_{21}$ when it is small. Therefore CP violation  can occur even if 
$\delta m^2_{12}=0$. However, CP violation is now possible with 
two neutrino oscillations. In addition, the CP effect with three neutrino
flavours, contrary to unitary oscillations, can be substantial even if
one of the elements of the mixing matrix is very small.

\section{CP-violating effects in neutrino oscillations}

CP violating  effects can be seen in appearence experiments.
Let us consider the
following standard quantities

\begin{eqnarray*}
  A_{CP}( \alpha ; \beta ) &  =& 
\frac{P(\nu_\alpha \to \nu_\beta) - P(\bar{\nu}_\alpha \to 
\bar{\nu}_\beta)}{P(\nu_\alpha \to \nu_\beta) + 
P(\bar{\nu}_\alpha \to 
\bar{\nu}_\beta)}, \\
A_{T} ( \alpha ; \beta ) & =& 
\frac{P(\nu_\alpha \to \nu_\beta) - P({\nu}_\beta \to 
{\nu}_\alpha)}{P(\nu_\alpha \to \nu_\beta) + P({\nu}_\beta \to 
{\nu}_\alpha)}.
\end{eqnarray*}

In vacuum $A_{CP}=A_T$. The same is true in the case of 
the new Eq.~\ref{prob} 
when a  nonunitary matrix U is present.

We assume the following values of the standard parameters

\begin{eqnarray}
\delta m_{21}^2 &=&5 \cdot 10^{-5}\;eV^2, \nonumber  \\ 
\delta m_{31}^2 &=&3 \cdot 10^{-3}\;eV^2,  \nonumber   \\ 
\Theta_{12} & \simeq & 35^o,\;
\Theta_{23}  \simeq  40^o,\; 
\Theta_{13}  \simeq  5^o, \label{smp} \\
\delta & \simeq & \pm 90^o. \nonumber  
\end{eqnarray}

These values are consistent with CHOOZ \cite{chooz},
the LMA MSW solution of the solar neutrino problem \cite{snp}
and the Superkamiokande data \cite{sk}.
For the nonstandard parameters we take 

\begin{eqnarray}
| \epsilon_1| &=& 0.001 \nonumber \\
| \epsilon_2| &=& 0.1 \label{nsp} \\
| \epsilon_3| &=& 0.1 \nonumber
\end{eqnarray}

which are consistent with Eqs.~\ref{ee},\ref{emu},\ref{mutau}.

\begin{figure}[h]
\epsfig{figure=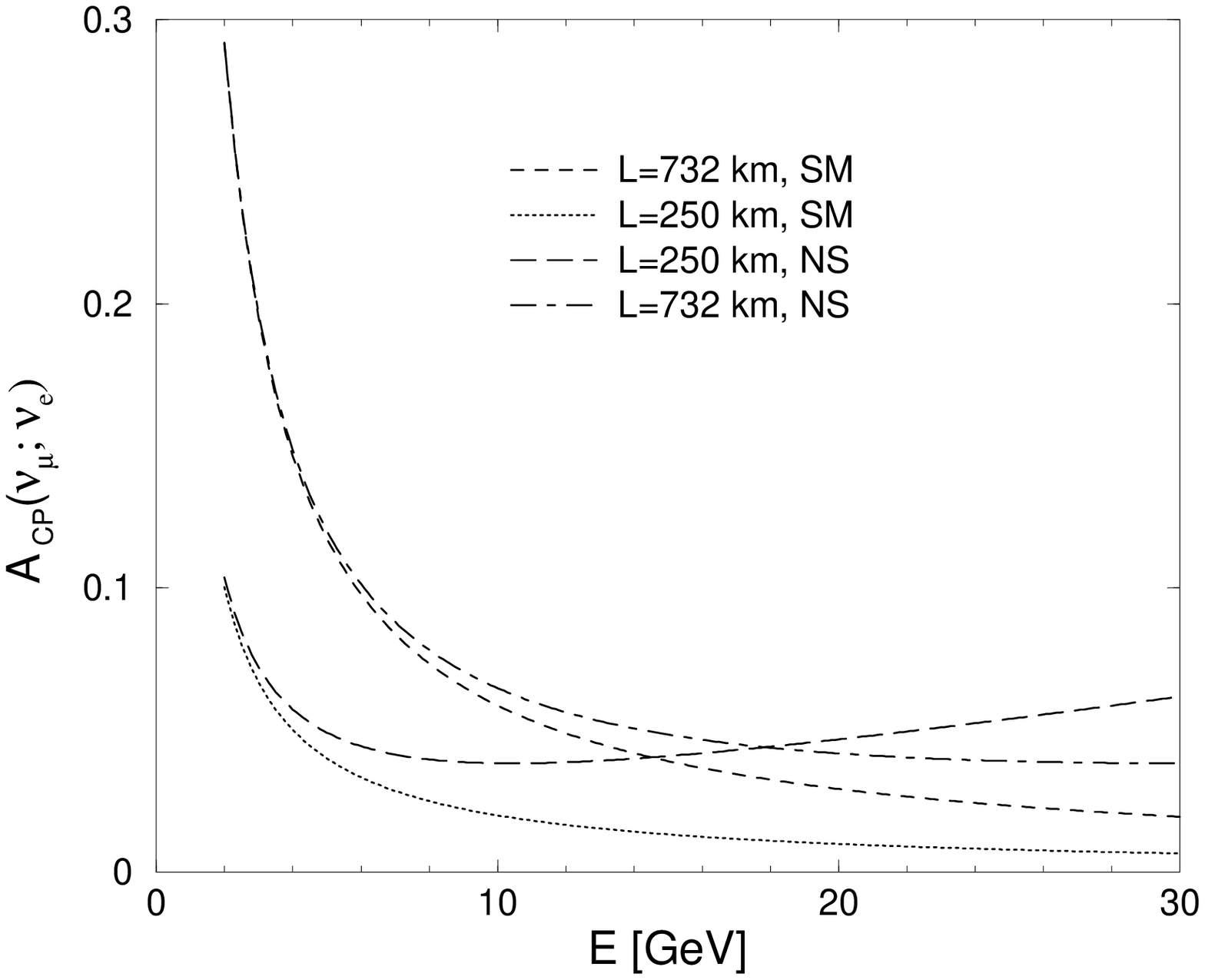, width=8.5cm}
\caption{The $A_{CP}( \mu ; e )$ asymmetry as function of
neutrino energy. The label 'NS' means that Eqs.~\ref{smp},\ref{nsp}
are taken into account. The label 'SM' means that the values of
neutrino parameters as given in Eq.~\ref{smp} and $\epsilon_i=0$
have been taken. 
\label{femu}}
\end{figure}

\begin{figure}[h]
\epsfig{figure=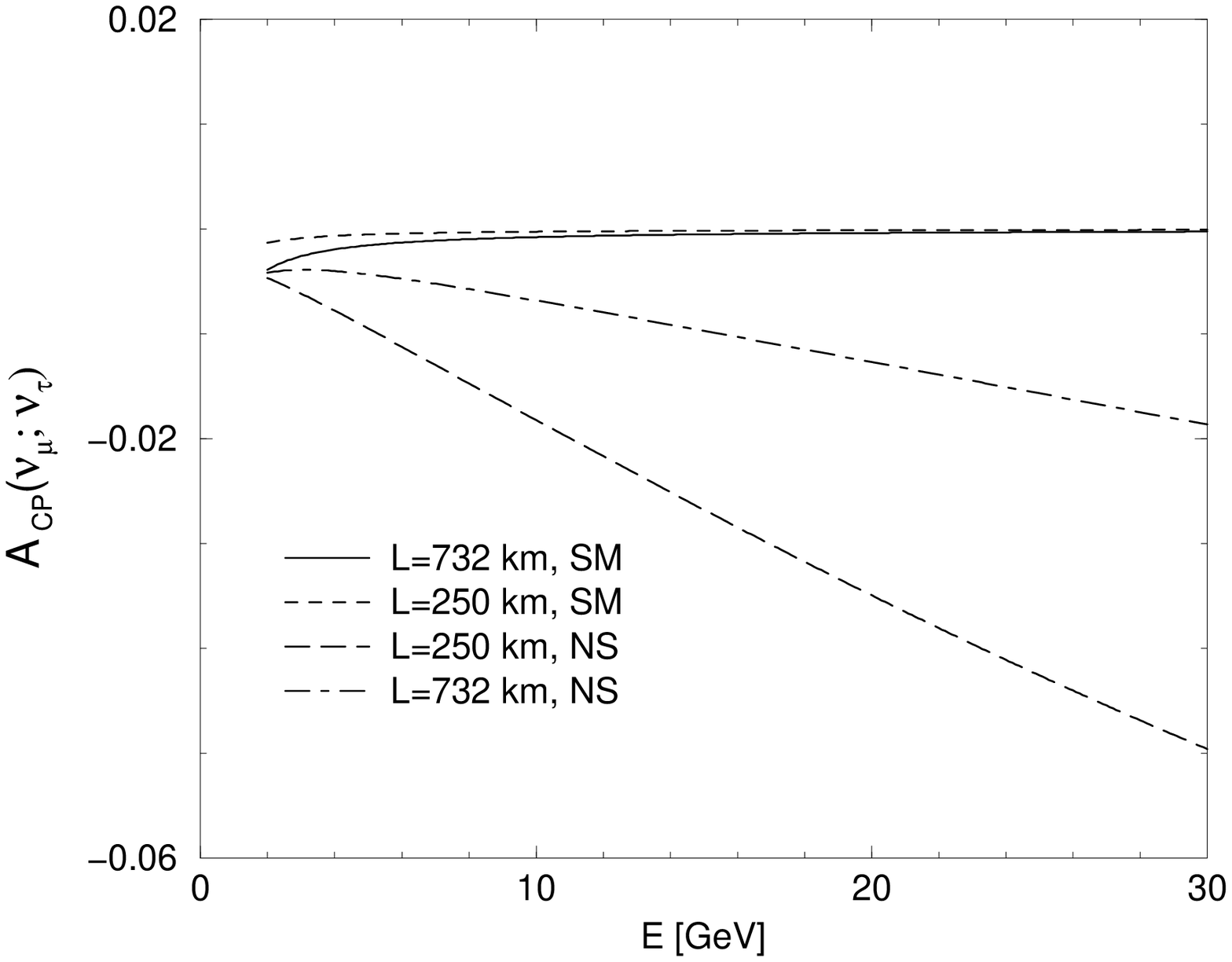, width=8.5cm}
\caption{The $A_{CP}( \mu ; \tau )$ asymmetry as function of
neutrino energy. The label 'NS' means that Eqs.~\ref{smp},\ref{nsp}
are taken into account. The label 'SM' means that the values of
neutrino parameters as given in Eq.~\ref{smp} and $\epsilon_i=0$
have been taken. 
\label{fmutau}}
\end{figure}

Figs.~\ref{femu},\ref{fmutau} show the results  
for  two cases of 
$A_{CP}( e ; \mu )$ and   $A_{CP}( \mu ; \tau )$,
and   
long-baseline (L=732 km, e.g MINOS)
or short-baseline (L=250 km, e.g. K2K) experiments. 
The neutrino energies are chosen to be
between 2 GeV and 30 GeV. 
We can see that the effects of the nonstandard heavy neutrino
mixings can be quite large, even much bigger than in the
unitary approach when $\epsilon_i=0$.

\begin{figure}[h]
\epsfig{figure=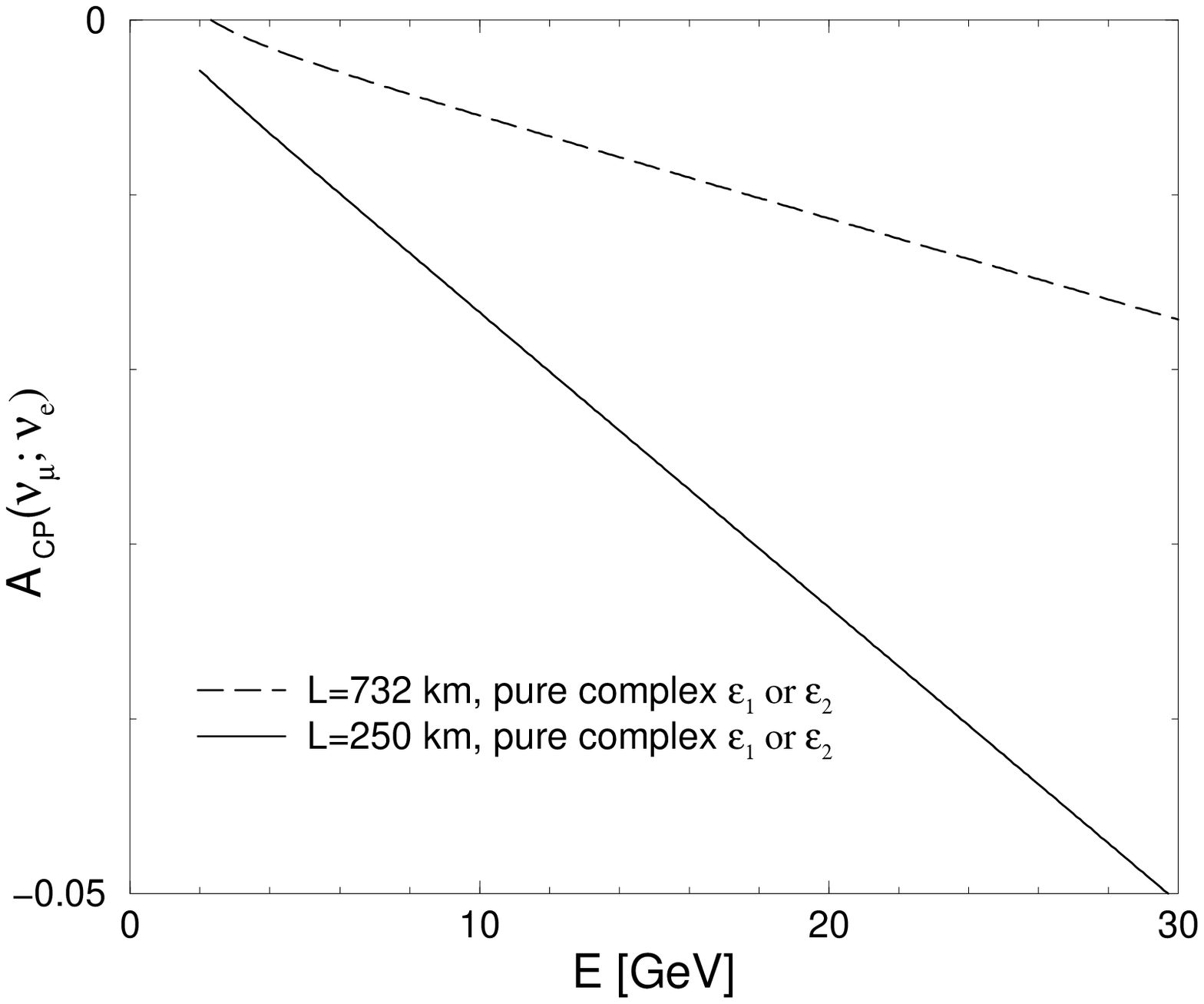, width=8.5cm}
\caption{The $A_{CP}( \mu ; e)$ asymmetry generated by the NS
sector. The results are for the parameters Eq.~\ref{smp}
but with $\delta=0$. 
$\epsilon_{1}=0.001 \cdot (1\; \mbox{\rm or}\; i)$, $\epsilon_{2}=0.1 \cdot (i\; \mbox{\rm or}\; 1)$,
$\epsilon_3=0.1$.
This choice is consistent with Eqs.~\ref{ee}-\ref{mutau}.
\label{fgemu}}
\end{figure}

\begin{figure}[h]
\epsfig{figure=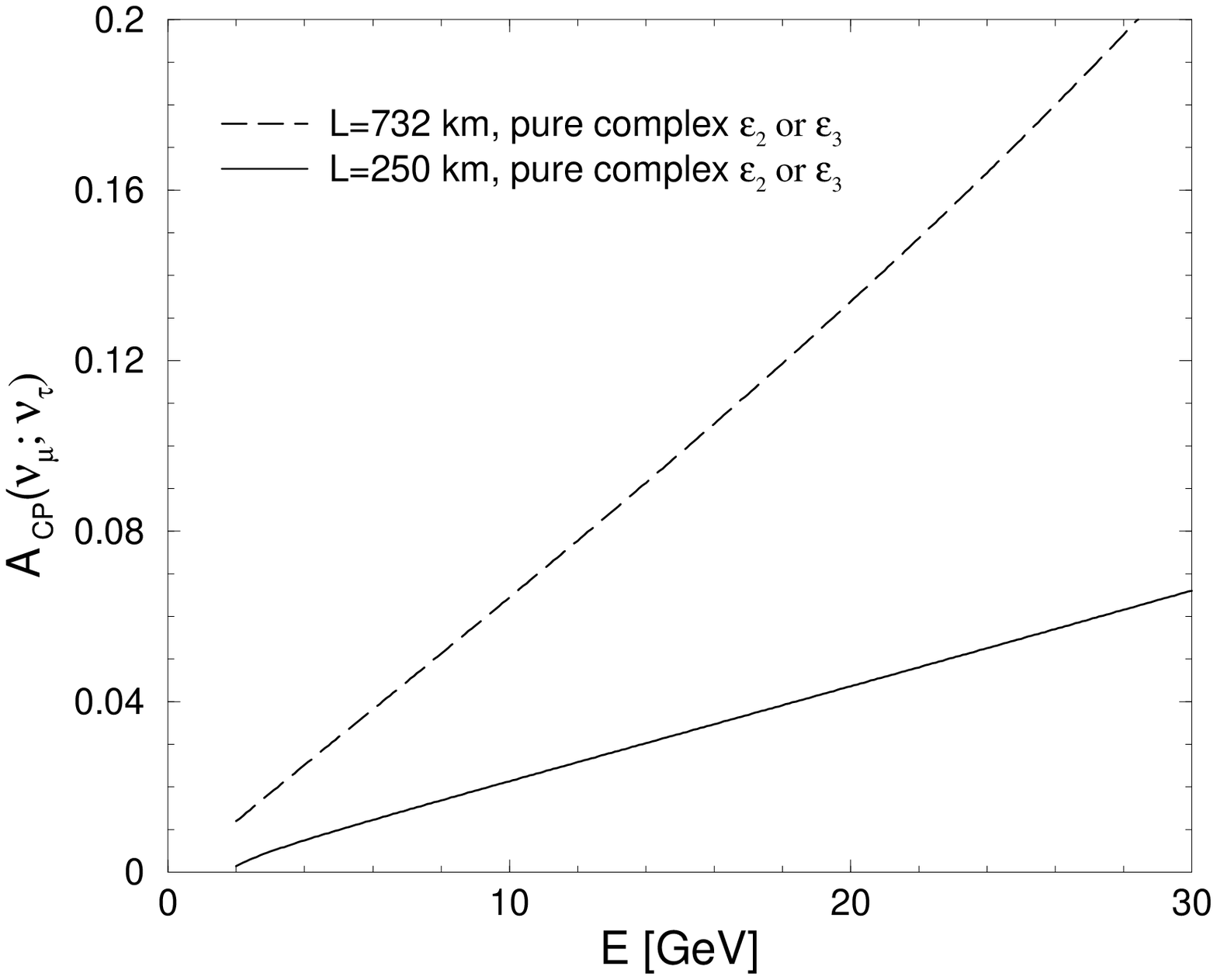, width=8.5cm}
\caption{The $A_{CP}( \mu ; \tau )$ asymmetry generated by the NS
sector. The results are for the parameters Eq.~\ref{smp}
but with $\delta=0$.  
$\epsilon_{2\; (\mbox{\rm or}\; 3)}=0.1 \cdot i$, $\epsilon_1=10^{-4}$.
This choice is consistent with Eqs.~\ref{ee}-\ref{mutau}.
\label{fgmutau}}
\end{figure}

In  Figs.~\ref{fgemu},\ref{fgmutau} the results are given
for genuine CP effects of NS sector when some of the $\epsilon_i$'s
are chosen to be complex and   $\delta=0$.

Three lessons can be learned  from  these numerical results. 
First of all short baseline experiments
are sensitive to the NS sector. Some improvements of the  
constraints Eqs.~\ref{ee},\ref{emu},\ref{mutau} are possible
in this case when no signal for $A_{CP}$ is found. Second,
the NS effects connected to the complexity of $\epsilon_i$ can mimic
SM effects of $\delta$.
Third, cancellations between the SM and NS effects can appear.

We would like to finish with a somehow academic example
of what more do
'nonorthogonal' neutrino states mean.

When a unitary U is used in the description of neutrino
oscillations, the following relation holds

\begin{equation}
\sum\limits_{\alpha} P_{\alpha \beta}=1,
\end{equation}

e.g.:
$$P_{ee}+P_{e \mu}+P_{e \tau}=1.$$

It simply means that the number of emitted neutrinos 
of the given flavour (e.g. e) will be the same as the number 
of final neutrinos of any type. 
However, for a nonunitary U  this relation is not 
fullfil. Let us see it in a simple case of two flavours,
when  U is defined as ($\Theta_2=\Theta_1+\epsilon$)

\begin{equation}
U=\left( \matrix{ \cos{\Theta_1} &  \sin{\Theta_1} \cr
                   -\sin{\Theta_2} &  \cos{\Theta_2}}
\right).
\end{equation}

In this case we get 
\begin{eqnarray}
\sum\limits_{\alpha=e,\mu} P_{e \alpha} & = &
P_{ee}+P_{e \mu}=1+4 \epsilon \sin^2{\Delta_{21}}  
\sin{\Theta_1} \cos{\Theta_1} \cos{2 \Theta_1} +
{\cal{O}} (\epsilon^2), \\
\sum\limits_{\alpha=e,\mu} P_{\mu \alpha} & = &
P_{\mu e}+P_{\mu \mu}=1-4 \epsilon \sin^2{\Delta_{21}}  
\sin{\Theta_1} \cos{\Theta_1} \cos{2 \Theta_1}
+{\cal{O}} (\epsilon^2).
\end{eqnarray}

We can see that the sum  can be either  larger or smaller than 
1. A similar result holds for a 3 dimensional U.

\end{document}